\documentclass[prl,twocolumn,superscriptaddress]{revtex4-1}

\usepackage[utf8]{inputenc}
\usepackage{amsmath,revsymb,amssymb}
\usepackage{graphicx}
\usepackage[colorinlistoftodos]{todonotes}
\usepackage{xcolor}
\usepackage{bbold}
\usepackage{braket,ulem}

\newcommand{\MS}{{\mathcal S}}
\newcommand{\Nab}{\mathcal N_{AB}}
\newcommand{\Uaj}{U_{A,j}}

\renewcommand{\emph}{\textit}

\begin{document}
    
    \title{    Direct Reconstruction of the Quantum Density Matrix by Strong Measurements}
    
    \author{Luca Calderaro}
    \affiliation{
        Dipartimento di Ingegneria dell'Informazione, Universit\`a di Padova, via Gradenigo 6B, 35131 Padova, Italy.}
    \affiliation{Centro di Ateneo di Studi e Attivit\`a Spaziali ``Giuseppe Colombo'',
        Universit\`a di Padova, via Venezia 15, 35131 Padova, Italy}
    \author{Giulio Foletto}
    \affiliation{
        Dipartimento di Ingegneria dell'Informazione, Universit\`a di Padova, via Gradenigo 6B, 35131 Padova, Italy.}
    \author{Daniele Dequal}
    \affiliation{Matera Laser Ranging Observatory, Agenzia Spaziale Italiana, Matera 75100, Italy}
    \author{Paolo Villoresi}
    \affiliation{
        Dipartimento di Ingegneria dell'Informazione, Universit\`a di Padova, via Gradenigo 6B, 35131 Padova, Italy.}
    \affiliation{
        Istituto di
        Fotonica e Nanotecnologie, CNR, via Trasea 7, 35131 Padova, Italy}
    \author{Giuseppe Vallone}
    \affiliation{
        Dipartimento di Ingegneria dell'Informazione, Universit\`a di Padova, via Gradenigo 6B, 35131 Padova, Italy.}
    \affiliation{
    	Istituto di
    	Fotonica e Nanotecnologie, CNR, via Trasea 7, 35131 Padova, Italy}

    \begin{abstract}
        New techniques based on weak measurements have recently been introduced to the field of quantum state reconstruction.
        Some of them allow the direct measurement of each matrix element of an unknown density operator and need
        only $O(d)$ different operations, compared to $d^2$ linearly independent projectors in the case of standard quantum state tomography, for the reconstruction of an arbitrary mixed state. 
        However, due to the weakness of these couplings, these protocols are approximated and prone to large statistical errors.
        We propose a method which is similar to the weak measurement protocols but works regardless of the coupling strength: our protocol is not approximated and thus improves the accuracy and precision of the results with respect to weak measurement schemes.
        We experimentally apply it to the polarization state of single photons and compare the results to those of preexisting methods for different values of the coupling strength. Our results show that our method outperforms previous proposals in terms of accuracy and statistical errors.
    \end{abstract}
    
    \maketitle
    
    \textit{Introduction}.\textemdash 
    The most common characterization of a general quantum state is given by a density operator and its determination
    is one of the most important problems of quantum mechanics.
    The usual way of reconstructing it is known as quantum state tomography (QST).
    If $d$ is the dimension of the system, QST employs
    $d^2$ {\it linearly independent}
    projectors~\cite{JamesQST,SchmiedQST}
    and can become impractical for large $d$.
    An alternative approach is based on the determination of the Moyal quasicharacteristic function by sequentially measuring two conjugate variables \cite{DiLo13prl,DiLo13pra}.
    
    Recently, Lundeen \textit{et al.} \cite{LundeenDWT, LundeenMixed} proposed new ways, called {\it direct reconstruction}, to determine the density matrix through weak measurements, processes in which the measuring device (called a pointer) perturbs only slightly the system which it is coupled to, so as to limit the collapse of its state \cite{AharonovSpin, DuckWeakValues, RitchieAmplification}. 
    These techniques have already been experimentally verified \cite{SalvailPolarization, ThekkadathDirectMeasurement} and thoroughly compared to QST \cite{MacconeComparison}.
    One such protocol can find an entire density matrix with only $d+1$ different unitary operations, a single $d$-outcome projective measurement and a small $d$-independent number of pointer measurements. Moreover, its steps involve only states in the measurement basis plus one off-basis component, as will be detailed below. However, the use of weak values implies that all these strategies are approximated and affected by great statistical errors: there is a trade-off between the validity of the approximation (improved by weakening the interaction) and the statistical uncertainties
    (improved by increasing the interaction).
    
    It has been shown that for pure states, it is possible to extend these schemes to arbitrary coupling strength (even to {\it strong measurements})
    without any approximation \cite{ValloneStrongMeasurements, zhang16pra, ZouStrong, denkmayr17prl,gross15pra}.
    Other methods valid for pure states were presented in Refs. \cite{Goyeneche15prl,Baldwin16pra}, whereas
    a generalization for mixed states was also proposed in Ref. \cite{ValloneStrongMeasurements} and in Ref. \cite{zhu16pra}.
    
    Here, we propose  
    a protocol for the (exact) direct reconstruction of the density operator without weak measurements.
    We also report on our experimental realization applied
    to the polarization state of single photons.
    We finally show that our method overcomes the weak measurement proposal \cite{ThekkadathDirectMeasurement} in terms of accuracy and statistical uncertainty, and we discuss the relation with standard QST.

    \textit{Theoretical model}.\textemdash 
    \begin{figure}
        \centering
        \vskip 0 pt
        \includegraphics[width=\columnwidth]{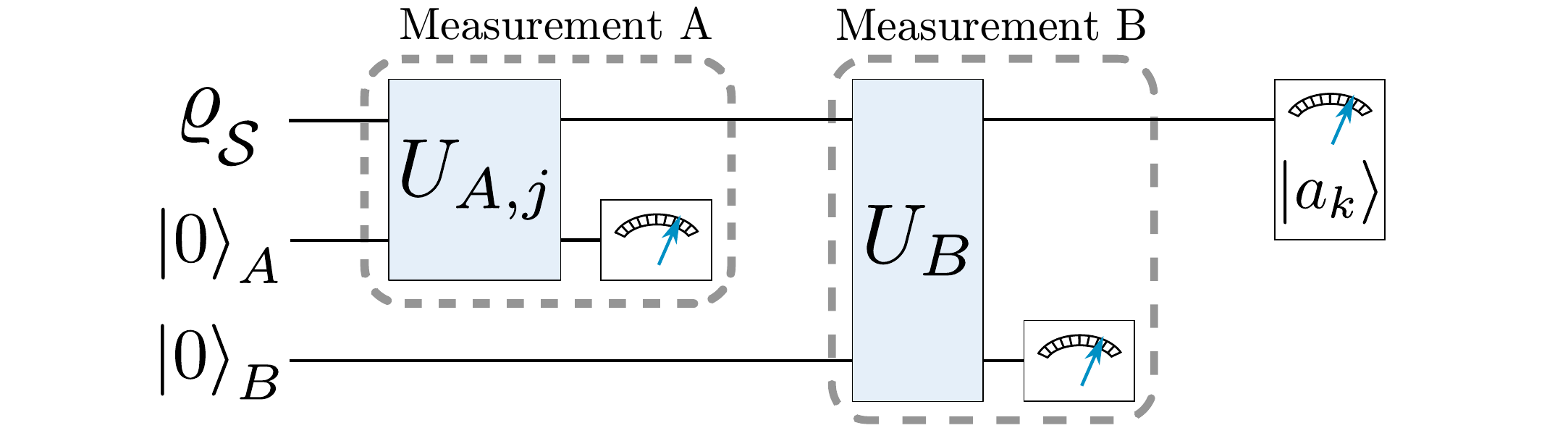}
        \caption{The reconstruction of the density matrix $\varrho_\MS$ needs two bidimensional pointers.
            Measurement $A$ ($B$) consists of a unitary evolution, $\Uaj$ ($U_B$), and a projective measurement on the pointer $A$	($B$). Finally, a projective measurement on the state $\ket{a_k}$ is performed.}
        \label{fig:Scheme}
    \end{figure}
    The weak measurement process uses a quantum pointer which represents the measuring device \cite{DresselWeakValues, SvenssonPedagogical}.
    We will focus on the case of bidimensional pointers, where the Pauli operators $X$, $Y$, $Z$ are naturally defined \cite{WuWeakQubit}.
    To weakly measure an observable $\mathcal O$ on a system $\mathcal S$, the pointer $A$ is initially prepared into $\ket{0}_A$, the $+1$ eigenstate of $Z$.
    Then, $\mathcal S$ and $A$ are coupled through the following unitary evolution:
    \begin{equation}
        U=e^{-i\theta \mathcal O\otimes Y}\,,
    \end{equation}
    where $\theta$ represents the strength of the coupling.
    In the weak approximation ($\theta \ll 1$), the expression for $U$ can be truncated at first order: 
    $U \approx \openone-i\theta \mathcal O \otimes Y $.
    However, when $\mathcal O=\Pi$ is a projector, 
    {the following equation holds exactly for any $\theta$:}
    \begin{equation}
        U=e^{-i\theta \Pi\otimes Y} = (\openone-\Pi) \otimes \openone + \Pi \otimes e^{-i\theta Y}\,.
        \label{eq:ExactEvolution}
    \end{equation}
    A value of $\theta=(\pi/2)$ represents a (maximally) strong measurement, because the two possible pointer states after the evolution are orthogonal.
    Then, a projection (called postselection) on a specific state of the system may occur.
    Finally, appropriate pointer measurements can extract information on the system.
    
    For the direct reconstruction of the density matrix $\varrho_\MS$, our protocol (and its weak counterpart \cite{LundeenMixed,ThekkadathDirectMeasurement}) needs two independent pointers $A$ and $B$ as illustrated in Fig.~\ref{fig:Scheme}.
    The initial system-pointer state is
    \begin{equation}
        \sigma_{in}=
        \varrho_\MS \otimes {\ket{0}}_A\bra{0}  \otimes {\ket{0}}_B\bra{0}\,.
    \end{equation}
    In order to express the $jk$th element $\varrho_{jk}$ of $\varrho_\MS$ in the orthonormal basis $\{\ket{a_j}\}$, $j=1,\ldots,d$ of the system Hilbert space, the projector $\Pi_{a_j}={\ket{a_j}}_\MS\bra{a_j}$ is coupled to the $Y_A$ operator of the first pointer:
    \begin{equation}
        \Uaj=e^{-i\theta_A \Pi_{a_j}\otimes Y_A} \otimes \openone_B\,.
        \label{UA}
    \end{equation}
    Then, the projector $\Pi_{b_0}={\ket{b_0}}_\MS\bra{b_0}$ is coupled to $Y_B$ of the other pointer, where $\ket{b_0}_\MS=\frac{1}{\sqrt{d}}  \sum_j \ket{a_j}_\MS$ independently of $j$ or $k$:
    \begin{equation}
        U_B=e^{-i\theta_B \Pi_{b_0}\otimes Y_B}\otimes \openone_A\,.
        \label{UB}
    \end{equation}
    These two evolutions must happen subsequently and strictly in this order, because $\Uaj$ and $U_B$ do not commute \cite{MitchinsonSequential,piacentini16prl}.
    After this, some useful measurements that we will detail below are performed on the pointers, while the system is (projectively) measured in the $\{\ket{a_k}\}$ basis.
    
    In the limit of weak couplings, it is possible to neglect all terms of order higher than $\theta_A\theta_B$ and find a ``weak'' (\textit{W}) estimate for the matrix element $\varrho_{jk}$ via the following pointer measurements \cite{LundeenMixed,ThekkadathDirectMeasurement}:
    \begin{equation}
        \begin{aligned}
            \text{Re}(\varrho^W_{jk}) &= \Nab  \left( \langle X_AX_B \rangle_{j,k} - \langle Y_AY_B\rangle_{j,k} \right) \\
            \text{Im}(\varrho^W_{jk}) &= \Nab \left( \langle Y_AX_B \rangle_{j,k} + \langle X_AY_B\rangle_{j,k}\right)\,,
        \end{aligned}
        \label{eq:lundeen}
    \end{equation}
    where we have defined the state-independent factor $\Nab=d/4\sin\theta_A\sin\theta_B$.
    This can be determined by knowing the values of $\theta_{A,B}$ or by normalizing $\varrho^W$. 
    We note that $\varrho^W$ correctly approximates the original density matrix $\varrho_\MS$ only for small $\theta_{A,B}$.
    For consistency with most of the literature, in the above equations we do not explicitly mention the projector $\Pi_{a_k}=\ket{a_k}_\MS\bra{a_k}$; however, the symbol $\langle X_AY_B\rangle_{j,k}$ indicates the mean value of observable $\Pi_{a_k} \otimes X_A \otimes Y_B$ on the tripartite state after the evolution $U_B \Uaj$, namely
    $\langle X_AY_B\rangle_{j,k} = \text{Tr}[\Pi_{a_k} \otimes X_A \otimes Y_B (U_B\Uaj\sigma_{in}\Uaj^\dag U^\dag_B)]$.
    We find that by using Eq.~\eqref{eq:ExactEvolution}, we can obtain an exact estimate of the density matrix (see the Supplemental Material \cite{SupMat}):
    \begin{equation}
        \begin{aligned}
            \text{Re} (\varrho^{I}_{jk}) =& \text{Re} (\varrho^W_{jk})+ 2\Nab (t_B\langle X_A\Pi_{1B} \rangle_{j,k} \\
            &+ t_A\langle \Pi_{1A}X_B \rangle_{j,k} + 2t_At_B\langle \Pi_{1A}\Pi_{1B} \rangle_{j,k}) \\
            \text{Im} (\varrho^{I}_{jk}) =& \text{Im} (\varrho^W_{jk}) + 2\Nab t_B\langle Y_A\Pi_{1B} \rangle_{j,k} 
        \end{aligned}
        \label{eq:Correct}
    \end{equation}
    where $\Pi_1=\ket1\bra1=(1-Z)/2$ is the projector on the $-1$ eingenstate of $Z$, and $t_{A,B}=\tan(\theta_{A,B}/2)$.
    This protocol, representing the direct generalization of the method proposed in Refs. \cite{LundeenMixed,ThekkadathDirectMeasurement}, needs more measurements than the weak counterpart of Eq. \eqref{eq:lundeen} (eight, rather than four, correlations should be measured for each element $\varrho_{jk}$). 
    However, as shown below, the increased complexity is compensated by a better estimation of the density matrix.
    Moreover, we also find a simpler expression, requiring even fewer measurements than Eq. \eqref{eq:lundeen}:
    \begin{equation}
        \begin{aligned}
            \varrho^{II}_{jj}&=16\Nab^2 \langle \Pi_{1A}\Pi_{1B} \rangle_{j,k}  &\quad	&\forall k \\
            \text{Re}(\varrho^{II}_{jk})&=-2\Nab  \langle Y_AY_B\rangle_{j,k}  &\quad	&j\neq k \\
            \text{Im}(\varrho^{II}_{jk})&=2\Nab\langle X_AY_B\rangle_{j,k}  &\quad	&j\neq k 
        \end{aligned}
        \label{eq:DRDO}
    \end{equation}
    Summarizing, these methods require the implementation of $d+1$ unitary operations ($\Uaj$ and $U_B$), one projective measurement on the $\{\ket{a_k}\}$ basis and a small number of pointer measurements [four observables for Eq.~\eqref{eq:lundeen}, eight for Eq.~\eqref{eq:Correct}, and three for Eq.~\eqref{eq:DRDO}].
    Moreover, they only need to select the $d$ components of the state of the system in the measurement basis plus the $\ket{b_0}$ component.
    
    The fundamental advantage of the latter two schemes is that they are accurate for any value of $\theta$. (From now on, we will only consider the case $\theta_A = \theta_B = \theta$ for simplicity, although these equations are also valid for $\theta_A \neq \theta_B$.) 
    Consequently, there is no need to tune the strength into the range of the weak approximation, and it is possible to use more practical strong measurements, which are more akin to standard quantum projections and less prone to statistical errors.
    Indeed, inverting the above relations shows that the experimental data are proportional to $\sin^2{\theta}$ [except for the diagonal elements in Eq. \eqref{eq:DRDO}, for which the factor is $\sin^4{\theta}$] that greatly weakens the signal when $\theta$ is small, making statistical errors more relevant.

    We can evaluate the \textit{mean square statistical error} $\delta \varrho$ as
    \begin{equation}
        \label{deltarho}
        \delta \varrho= \sqrt{\sum_{j,k}|\delta\varrho_{jk}|^2} \,.
    \end{equation}
    If $N$ labels the number of events used to measure each correlation term in Eq. \eqref{eq:lundeen} [or in Eqs. \eqref{eq:Correct} and \eqref{eq:DRDO}], we find that in the weak approximation, $\delta\varrho$ has a lower bound:
    \begin{equation}
        \delta \varrho \geq \frac{\alpha(d)}{\theta^2 \sqrt{N} }\,,
        \label{deltarho_bound}
    \end{equation}
    where $\alpha(d)= [(d-1)\sqrt{d}]/[2\sqrt{2}]$ for the protocols of Eqs.~\eqref{eq:lundeen} and \eqref{eq:Correct} and $\alpha(d)= [\sqrt{d(d-1)(d-4)}]/2$ for Eq.~\eqref{eq:DRDO}. These equations were derived under 
    the assumption of reconstructing the entire matrix, taking its Hermitian part, and normalizing it.
    The factor $1/\theta^2$ highlights that the weaker the coupling is, the greater the statistical errors become.
    
    Weak measurements are also more vulnerable to  experimental systematic biases.
    Indeed, terms like $\langle X_AY_B \rangle_{j,k} $ are obtained from linear combinations of four contributions, each of them representing one projector that appears in the spectral decomposition of the Pauli operators.
    A small proportional bias in one of them becomes relevant when such linear combinations are also bound to be small by factor $\sin^2{\theta}$.
    
    In conclusion, our methods break the aforementioned trade-off of weak measurements. 
    They work with no approximation in the strong measurement regime, which allows more precise and accurate results.
    Moreover, the method of Eq. \eqref{eq:DRDO} needs fewer measurements than its weak counterpart (three different pointer operators, $X_AY_B$, $Y_AY_B$, $\Pi_{1A}\Pi_{1B}$, instead of four).

    \begin{figure}[t]
        \centering
        \vskip 0 pt
        \includegraphics[width=\linewidth,clip]{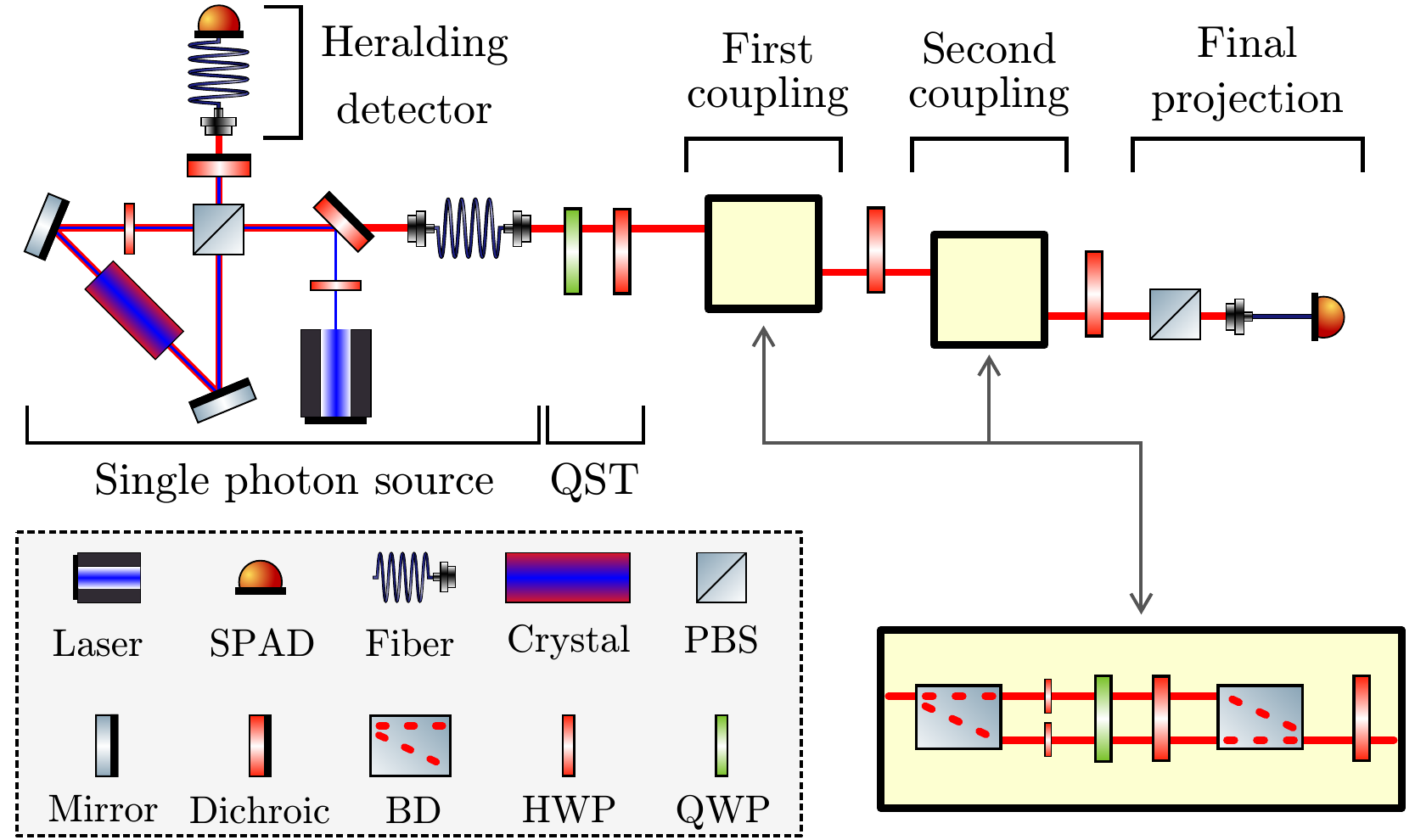}
        \caption{Experimental scheme.}
        \label{fig:ExpScheme}
    \end{figure}
    \textit{Experimental implementation}.\textemdash
    The experiment is shown in Fig.~\ref{fig:ExpScheme}.
    We produce polarization-entangled photon pairs with 
    a 30 mm periodically poled KTP crystal in a Sagnac interferometer \cite{kim06pra,schiavon16SciRep,schiavon17qst}.
    Degenerate down-converted photons at $809$~nm are collected into two single-mode fibers.
    One photon 
    is used as a trigger,
    while the other is sent to the measurement apparatus, which reconstructs its polarization state.
    We can adjust the purity of the measured qubit 
    using the half-wave plate (HWP) before the Sagnac interferometer.

    The basic block in our implementation of the unitary evolutions in Eq. \eqref{eq:ExactEvolution} is a Mach-Zehnder interferometer.
    A beam displacer (BD) is placed so that the two orthogonal components $\ket{a_0}$ and $\ket{a_1}$ of the photon polarization are spatially separated. The quantum state in now encoded in the photon path, 
    with polarization as the pointer.
    The strength of the coupling 
    is controlled by two twin HWPs, one for each arm of the interferometer.
    When $S=\Pi_{a_0}$, the polarization (pointer) is rotated about an angle $\theta$ by the HWP on the path corresponding to the $\ket{a_0}$ component. 
    This is achieved by aligning its optical axis at an angle $\theta/2$ compared to $\ket{a_0}$.
    To realize the unitary in the case of $S=\Pi_{a_1}$, we swap the role of the two paths by adding a HWP before the initial BD.
    The measurement of $X$ and $Y$ on the pointer state is performed using a quarter-wave Plate (QWP) and a HWP, whereas the measurement of $Z$ is done by blocking alternately one arm of the interferometer.
    Each configuration of these plates and of the shutters corresponds to a particular pointer projector that appears in the spectral decomposition of one Pauli operator.
    A second BD, which is oriented like the first one, closes the interferometer, and a HWP with axis at $45^\circ$ relative to $\ket{a_0}$ is placed after it.
    Only one of the two exit paths is considered: here, the polarization is the result of the collapse of the system after a successful pointer projection identified by the orientation of the internal plates.
    \begin{figure}[t]
        \centering
        \vskip 0 pt
        \includegraphics[width=0.95\columnwidth]{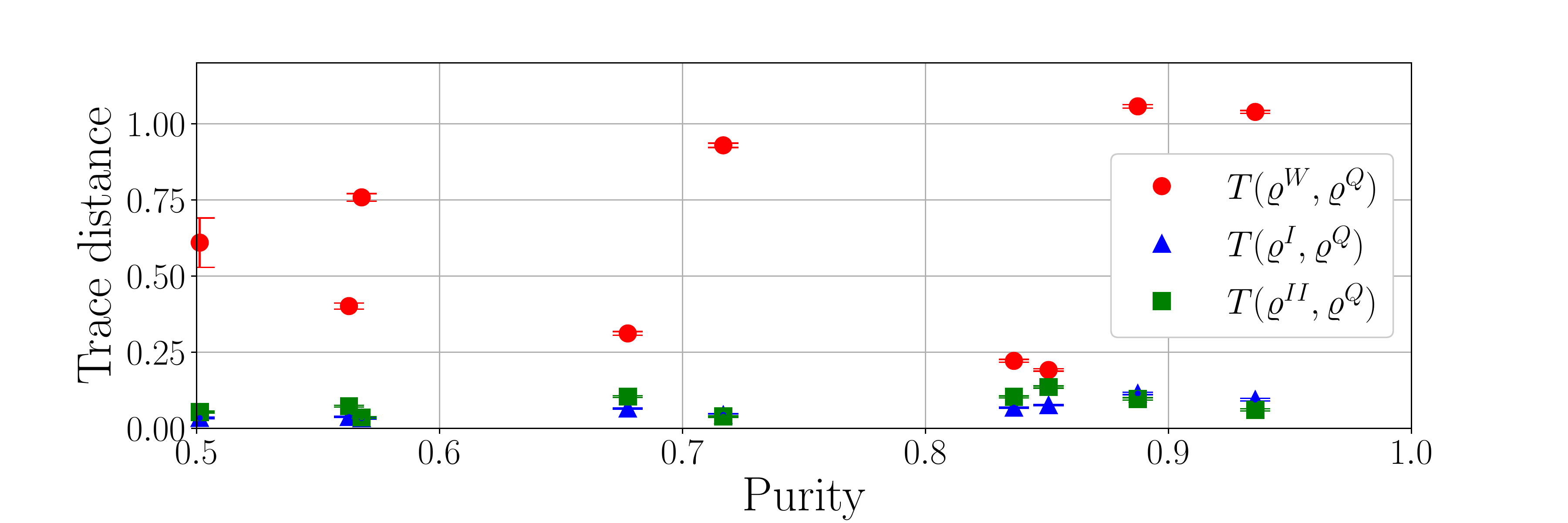}
        \caption{Trace distance between reconstructed states and reference states for different input purity: $T(\varrho^W, \varrho^{Q})$ (circles), $T(\varrho^{I}, \varrho^{Q})$ (triangles), $T(\varrho^{II}, \varrho^{Q})$ (squares).}
        \label{fig:TraceVsPurity}
    \end{figure}
    In our apparatus, we placed two interferometers in cascade, implementing $\Uaj$ and $U_B$.
    Since we chose $\ket{a_0}$ and $\ket{a_1}$ to be the horizontal $\ket{H}$ and vertical $\ket{V}$ polarizations respectively, we have that $\ket{b_0}$ is the diagonal polarization $\ket{b_0}=\ket{D}=(\ket{H} + \ket{V})/\sqrt{2}$.
    For this reason, two HWPs at $22.5^\circ$ are placed one in front and one at the end of the second interferometer.
    Finally, a polarizer projects the system on $\ket{a_k}$. 
    
    A small change in the scheme allows us also to perform the standard QST.
    By placing an additional QWP before the first HWP and blocking one arm of the second interferometer, we can project the polarization of the photon on the states $\ket{H}$, $\ket{V}$, $\ket{D}$, and $\ket{R} = (\ket{H}-i\ket{V})/\sqrt{2}$, thus obtaining the density matrix $\varrho^Q$ \cite{JamesQST}.
    
    \textit{Results}.\textemdash
    We chose the  trace distance to compare the reconstructed states given by Eqs. \eqref{eq:lundeen}, \eqref{eq:Correct}, and \eqref{eq:DRDO} with
    $\varrho^Q$, which is used as a fixed reference for all values of the coupling strength.
    For a fair comparison, we kept the reconstructed states as raw as possible by taking their Hermitian part and normalizing the trace, but without applying any postprocessing  such as a maximum likelihood estimation (MLE)~\cite{JamesQST}.
    
    \begin{figure}
        \centering
        \vskip 0 pt
        \includegraphics[width=0.99\columnwidth]{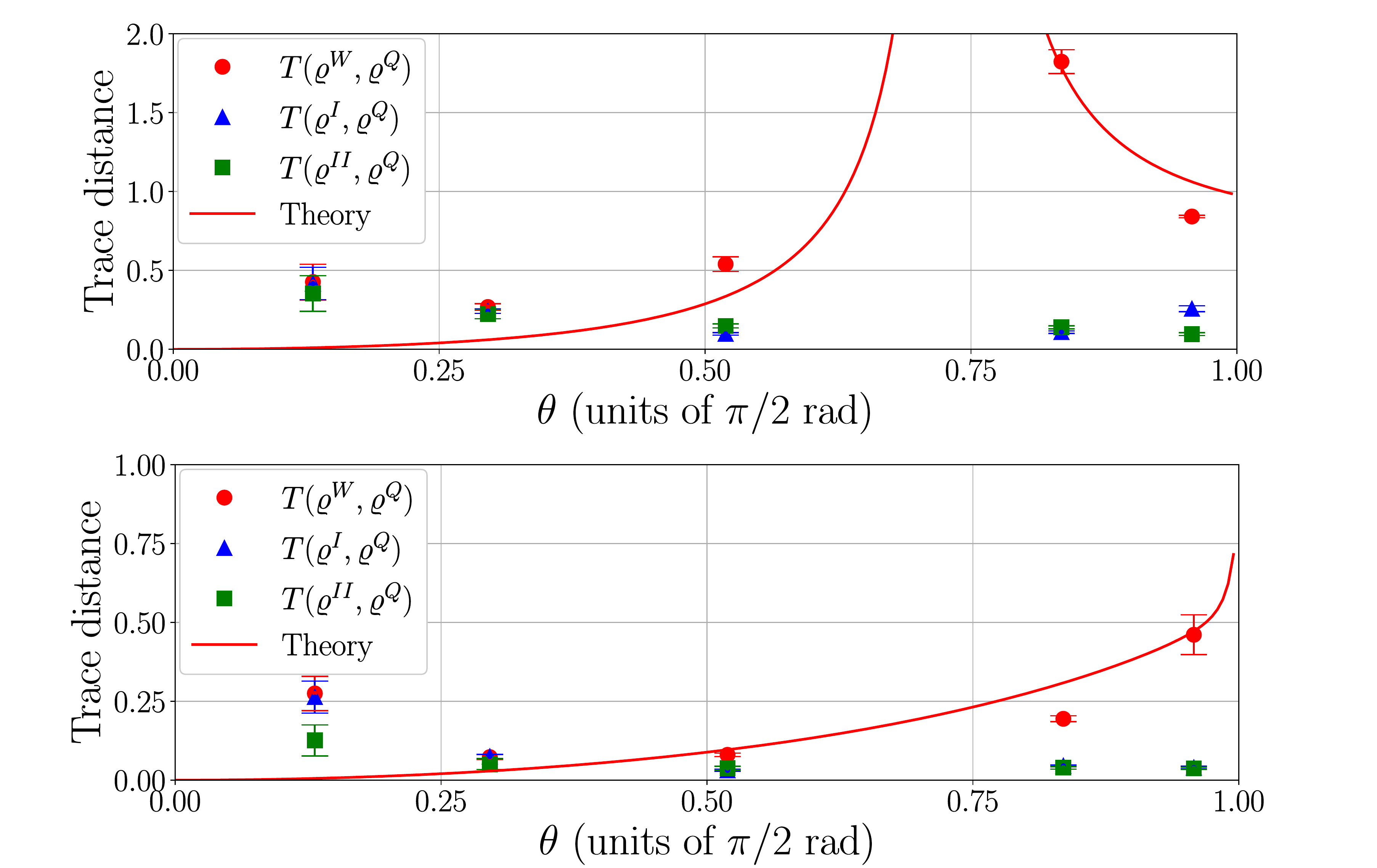}
        \caption{Trace distance between reconstructed states and reference state, for the input states (top) $\ket{D}$ and (bottom) maximally mixed.
            The solid line represents the theoretical trace distance between the expected value of $\varrho^W$ and the experimental $\varrho^Q$.}
        \label{fig:Tracedistances}
    \end{figure}
    
    Using the HWP in our source, we generated nine states of different purities, measured as ${\rm Tr}[(\varrho^Q)^2]$.
    We applied to them the three methods at maximum strength ($\theta=\pi/2$).
    Figure~\ref{fig:TraceVsPurity} reports the trace distance between the measured reconstructed states and the reference state: $T(\varrho^W, \varrho^{Q})$ (circles), $T(\varrho^{I}, \varrho^{Q})$ (triangles), $T(\varrho^{II}, \varrho^{Q})$ (squares).
    Predictably, the high value of strength invalidates the weak approximation and makes the results of the method in Eq. \eqref{eq:lundeen} inaccurate.
    There is no direct relation between purity and these errors, which are rather influenced by the particular input states.
    Instead, our two protocols show a lower trace distance in the entire range of purity, confirming the validity of our proposals for both mixed and almost pure states.
    
    In the second phase of the experiment, we varied the coupling strength and focused only on two fixed input states.
    The results are shown in Fig.~\ref{fig:Tracedistances} with the two prepared states being $\ket{D}$ (top) and the maximally mixed one (bottom).
    The solid lines display the expected trace distance for the calculated state $\varrho^W$ of Eq.~\eqref{eq:lundeen} using the experimental $\varrho^{Q}$ as an estimate of $\varrho_\MS$.
    Again, they show that the weak approach is not reliable 
    as the strength increases.
    In the strong regime, these curves well reproduce the measured data $T(\varrho^W, \varrho^{Q})$, but we see larger trace distances than expected for weak coupling.
    This is probably due to the aforementioned greater vulnerability to small inaccuracies in the pointer projections, which only become relevant at low values of $\theta$.
    The protocol of Eq.~\eqref{eq:Correct} is similarly affected, because $\varrho^{I} \approx \varrho^W$ for small strength as the higher-order corrective terms become negligible.
    These biases have the same effect in the extraction of nondiagonal elements of Eq.~\eqref{eq:DRDO}, but terms like $\langle \Pi_{1A}\Pi_{1B} \rangle_{j,k}$ in the diagonal elements are a source of errors too.
    Indeed, the corresponding photon counts are weakened by a factor of $\sin^4{\theta}$, which is extremely small for low strength.
    Errors in the experimental realization of the projector or miscalculations of $\theta$ can cause the diagonal matrix elements to be much greater than expected, and the subsequent normalization of the trace can render the nondiagonal ones close to zero.
    This explains the high trace distance in Fig.~\ref{fig:Tracedistances}~(top) for $\varrho_\MS =\ket{D}\bra{D}$ (for which the nondiagonal elements are 0.5) and the slightly lower one in Fig.~\ref{fig:Tracedistances}~(bottom) for the maximally mixed state (which has null nondiagonal elements).
    
    \begin{figure}
        \centering
        \vskip 0 pt
        \includegraphics[width=0.99\columnwidth]{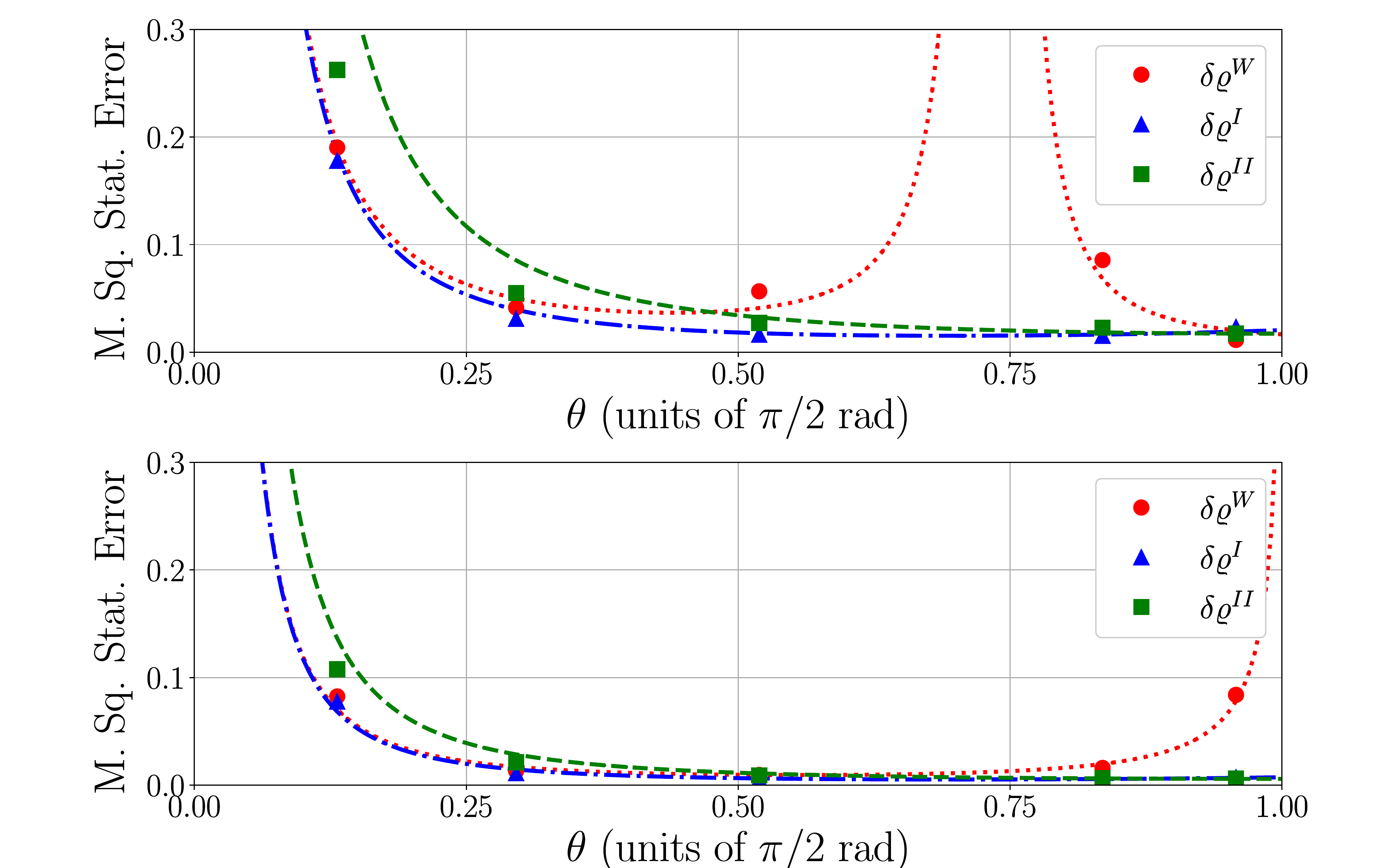}
        \caption{Mean square statistical error on the reconstructed states, for the input states (top) $\ket{D}$ and (bottom) maximally mixed.	
            The lines represent the theoretical expectations for $\delta\varrho^W$ (points), $\delta\varrho^I$ (dashes and points), and $\delta\varrho^{II}$ (dashes).}
        \label{fig:AbsErrors}
    \end{figure}
    
    However, Fig.~\ref{fig:Tracedistances} also shows how $\varrho^{I}$ and $\varrho^{II}$ become compatible with the reference state for large values of strength, confirming the correctness of these approaches.
    It is also clear that the weak measurement proposal is not accurate even at very small $\theta$ due to the high sensitivity to imperfections and systematic errors. Our method presents similar features at low $\theta$, but it drastically improves the performance at large $\theta$.
    
    We also evaluated the mean square statistical error $\delta\varrho$ given in Eq.~\eqref{deltarho} associated with the reconstructed matrices, which can be seen in Fig.~\ref{fig:AbsErrors}.
    The lines display the theoretical expectation values of these errors,  evaluated with
    the total number of events $N\approx 8 \times 10^3$ and $N \approx 4 \times 10^4$ 
    used in the experiments (top and bottom, respectively).
    Our results closely follow such curves and further prove that the errors dramatically increase for small values of $\theta$.
    The lower trace distances and  statistical errors on the right sides of Figs.~\ref{fig:Tracedistances} and \ref{fig:AbsErrors} clearly demonstrate the superiority of our strong measurement method with respect to the weak counterparts.
    Increasing the value of $\theta$, as allowed by the method proposed here, has a double advantage: it reduces the statistical errors and makes the protocol more robust against imperfections.
    
    \textit{Discussion}.\textemdash
    We have proposed a scheme 
    to directly reconstruct the density matrix that extends the existing idea based on weak measurement, making it exact for any value of the coupling strength with the pointers.
    In particular, we have shown that the use of strong measurements makes our protocol less vulnerable to experimental statistical and systematic errors in comparison to the original proposal~\cite{ThekkadathDirectMeasurement}, while the lack of approximations in our expressions makes the results devoid of any inherent biases. 
    
    In particular, our method uses {\it the same resources and experimental operations} of the weak counterpart, but it achieves much better performance.
    
    It is worth noticing that our protocol, the weak measurement proposal~\cite{ThekkadathDirectMeasurement}, or the QST \cite{JamesQST}
    can extract a raw density matrix from the system. As also underlined in Ref. \cite{ThekkadathDirectMeasurement}, if (semi)positivity is required, a postprocessing on the data (such as MLE or equivalent techniques) is necessary {\it for any method}. 
    On the other hand, if a single matrix element is required (in this case semipositivity cannot be verified), Eqs.~\eqref{eq:Correct} and \eqref{eq:DRDO} give it directly and without approximation in terms of the measured observables.
    As for any measurement, the single matrix element will be affected by experimental errors that can be mitigated by measuring all matrix elements and imposing positivity.
    We note that in the QST framework, the single matrix element $\varrho_{jk}$ can be obtained by measuring the four projectors $\Pi_{a_j}$, $\Pi_{a_k}$, $\Pi_{\ket{+_{jk}}}$, and $\Pi_{\ket {i_{jk}}}$ with $\ket{+_{jk}}=2^{-1/2}(\ket{a_j}+\ket{a_k})$
    and $\ket {i_{jk}}=2^{-1/2}(\ket{a_j}+i\ket{a_k})$. 
    However, differently from the direct method, this approach requires projectors on states that are outside of the basis used to express the density matrix and that are different for each matrix element. If all elements are required, $d^2$ independent projectors are needed, and for large dimension
    systems this may become a very hard experimental challenge.
    
    For this reason, we advocate that our scheme might be preferable to QST when the dimension $d$ of the system is large.
    Indeed, to determine all the matrix elements, it is sufficient to realize $d+1$ unitary operations $\Uaj$ and $U_B$, one $d$-outcome projective measurement on the system, and a limited number of pointer measurements [just three in the case of Eq. \eqref{eq:DRDO}]. All of these only involve states of the measurement basis plus one off-basis component ($\ket{b_0}$), a beneficial feature especially for systems that have a clearly preferred basis which is experimentally more accessible.

    Our experimental realization proves the validity of our proposal and shows that strong measurements are a feasible and convenient way to reconstruct the density operator even in the single-photon regime.
    
    \begin{acknowledgments}
        We acknowledge Dr. Matteo Schiavon for the useful discussions on the entangled photons source. 
        L. C. and G. F. contributed equally to this work.
    \end{acknowledgments}
    
    %

\begin{thebibliography}{10}
        
        \bibitem{JamesQST}
        D.~F.~V. James, P.~G. Kwiat, W.~J. Munro, and A.~G. White, Measurement of
            qubits, Phys. Rev. A \textbf{64}, 052312 (2001).
        
        \bibitem{SchmiedQST}
        R.~Schmied, Quantum state tomography of a single qubit: comparison of
            methods, J. Mod. Opt. \textbf{63}, 1744 (2016).
        
        \bibitem{DiLo13prl}
        A.~Di~Lorenzo, Sequential measurement of conjugate variables as an
            alternative quantum state tomography, Phys. Rev. Lett. \textbf{110}, 010404
        (2013).
        
        \bibitem{DiLo13pra}
        A.~Di~Lorenzo, Quantum state tomography from a sequential measurement of
            two variables in a single setup, Phys. Rev. A \textbf{88}, 042114 (2013).
        
        \bibitem{LundeenDWT}
        J.~S. Lundeen, B.~Sutherland, A.~Patel, C.~Stewart, and C.~Bamber, Direct
            measurement of the quantum wavefunction, Nature (London) \textbf{474}, 188 (2011).
        
        \bibitem{LundeenMixed}
        J.~S. Lundeen and C.~Bamber, Procedure for direct measurement of general
            quantum states using weak measurement, Phys. Rev. Lett. \textbf{108}, 070402
        (2012).
        
        \bibitem{AharonovSpin}
        Y.~Aharonov, D.~Z. Albert, and L.~Vaidman, How the result of a
            measurement of a component of the spin of a spin-1/2 particle can turn out to
            be 100, Phys. Rev. Lett. \textbf{60}, 1351 (1988).
        
        \bibitem{DuckWeakValues}
        I.~M. Duck, P.~M. Stevenson, and E.~C.~G. Sudarshan, The sense in which a
            "weak measurement" of a spin-1/2 particle's spin component yields a value
            100, Phys. Rev. D \textbf{40}, 2112 (1989).
        
        \bibitem{RitchieAmplification}
        N.~W.~M. Ritchie, J.~G. Story, and R.~G. Hulet, Realization of a
            measurement of a ``weak value'', Phys. Rev. Lett. \textbf{66}, 1107 (1991).
        
        \bibitem{SalvailPolarization}
        J.~Z. Salvail, M.~Agnew, J.~S. Allan, E.~Bolduc, J.~Leach, and R.~W. Boyd,
        Full characterization of polarization states of light via direct
            measurement, Nat. Photonics \textbf{7}, 316 (2013).
        
        \bibitem{ThekkadathDirectMeasurement}
        G.~S. Thekkadath, L.~Giner, Y.~Chalich, M.~J. Horton, J.~Banker, and J.~S.
        Lundeen, Direct measurement of the density matrix of a quantum system,
        Phys. Rev. Lett. \textbf{117}, 120401 (2016).
        
        \bibitem{MacconeComparison}
        L.~Maccone and C.~C. Rusconi, State estimation: A comparison between
            direct state measurement and tomography, Phys. Rev. A \textbf{89}, 022122
        (2014).
        
        \bibitem{ValloneStrongMeasurements}
        G.~Vallone and D.~Dequal, Strong measurements give a better direct
            measurement of the quantum wave function, Phys. Rev. Lett. \textbf{116},
        040502 (2016).
        
        \bibitem{zhang16pra}
        Y.-X. Zhang, S.~Wu, and Z.-B. Chen, Coupling-deformed pointer observables
            and weak values, Phys. Rev. A \textbf{93}, 032128 (2016).
        
        \bibitem{ZouStrong}
        P.~Zou, Z.~Zhang, and W.~Song, Direct measurement of general quantum
            states using strong measurement, Phys. Rev. A \textbf{91}, 052109 (2015).
        
        \bibitem{denkmayr17prl}
        T.~Denkmayr, \emph{et~al.}, Experimental demonstration of direct path
            state characterization by strongly measuring weak values in a matter-wave
            interferometer, Phys. Rev. Lett. \textbf{118}, 010402 (2017).
        
        \bibitem{gross15pra}
        J.~A. Gross, N.~Dangniam, C.~Ferrie, and C.~M. Caves, Novelty, efficacy,
            and significance of weak measurements for quantum tomography, Phys. Rev. A
        \textbf{92}, 062133 (2015).
        
        \bibitem{Goyeneche15prl}
        D.~Goyeneche, \emph{et~al.}, Five measurement bases determine pure
            quantum states on any dimension, Phys. Rev. Lett. \textbf{115}, 090401
        (2015).
        
        \bibitem{Baldwin16pra}
        C.~H. Baldwin, I.~H. Deutsch, and A.~Kalev, Strictly-complete
            measurements for bounded-rank quantum-state tomography, Phys. Rev. A
        \textbf{93}, 052105 (2016).
        
        \bibitem{zhu16pra}
        X.~Zhu, Y.-X. Zhang, and S.~Wu, Direct state reconstruction with
            coupling-deformed pointer observables, Phys. Rev. A \textbf{93}, 062304
        (2016).
        
        \bibitem{DresselWeakValues}
        J.~Dressel, M.~Malik, F.~M. Miatto, A.~N. Jordan, and R.~W. Boyd,
        Understanding quantum weak values: Basics and applications, Rev. Mod.
        Phys. \textbf{86}, 307 (2014).
        
        \bibitem{SvenssonPedagogical}
        B.~Svensson, Pedagogical review of quantum measurement theory with an
            emphasis on weak measurements, Quanta \textbf{2}, 18 (2013).
        
        \bibitem{WuWeakQubit}
        S.~Wu and K.~Mølmer, Weak measurements with a qubit meter, Phys.
        Lett. A \textbf{374}, 34  (2009).
        
        \bibitem{MitchinsonSequential}
        G.~Mitchison, R.~Jozsa, and S.~Popescu, Sequential weak measurement,
        Phys. Rev. A \textbf{76}, 062105 (2007).
        
        \bibitem{piacentini16prl}
        F.~Piacentini \emph{et~al.}, Measuring incompatible observables by
            exploiting sequential weak values, Phys. Rev. Lett. \textbf{117}, 170402
        (2016).
        
        \bibitem{SupMat}
        See Supplemental Material below for a proof of our results.
        
        \bibitem{kim06pra}
        T.~Kim, M.~Fiorentino, and F.~N.~C. Wong, Phase-stable source of
                polarization-entangled photons using a polarization Sagnac interferometer,
        Phys. Rev. A \textbf{73}, 012316 (2006).
        
        \bibitem{schiavon16SciRep}
        M.~Schiavon, G.~Vallone, and P.~Villoresi, Experimental realization of
            equiangular three-state quantum key distribution, Sci. Rep.
        \textbf{6}, 30089 (2016).
        
        \bibitem{schiavon17qst}
        M.~Schiavon, L.~Calderaro, M.~Pittaluga, G.~Vallone, and P.~Villoresi,
        Three-observer bell inequality violation on a two-qubit entangled
            state, Quantum Sci. Technol. \textbf{2}, 015010 (2017).
        
    \end{thebibliography}

    \clearpage
    \appendix
    \onecolumngrid
    \section{Supplemental Material: Demonstration of the main results}
    
    The initial system-pointer state is:
    \begin{equation}
    \sigma_{in}=
    \varrho_\MS \otimes {\ket{0}}_A\bra{0}  \otimes {\ket{0}}_B\bra{0}\,.
    \end{equation}
    The couplings with the pointers
    are expressed by the evolution operators
    \begin{equation}
    \Uaj=e^{-i\theta_A \Pi_{a_j}\otimes Y_A} \otimes \openone_B=
    (\openone_\MS-\Pi_{a_j})\otimes \openone_A\otimes\openone_B+
    \Pi_{a_j}\otimes e^{-i\theta_A Y_A} \otimes \openone_B
    \,.
    \end{equation}
    and
    \begin{equation}
    U_B=e^{-i\theta_B \Pi_{b_0}\otimes Y_B}\otimes \openone_A
    =(\openone_\MS-\Pi_{b_0})\otimes
    \openone_A\otimes\openone_B+
    \Pi_{b_0}\otimes\openone_A\otimes
    e^{-i\theta_B Y_B}\,.
    \end{equation}
    where $\Pi_{a_j}={\ket{a_j}}_\MS\bra{a_j}$ and
    $\Pi_{b_0}={\ket{b_0}}_\MS\bra{b_0}$ where $\ket{b_0}_\MS=\frac{1}{\sqrt{d}}  \sum_j \ket{a_j}_\MS$.

    
    The state after the interaction becomes
    \begin{equation}
    \begin{aligned}
    \sigma_{out,j}&=
    U_B\Uaj\sigma_{in}\Uaj^\dag U^\dag_B
    \end{aligned}
    \end{equation}
    We now evaluate explicitly the terms appearing in eq. (7) and (8)
    of the main text. We recall that we indicate by 
    $\langle O_AO_B\rangle_{j,k}$ the following expectation value:
    \begin{equation}
    \langle O_AO_B\rangle_{j,k} = \text{Tr}[(\Pi_{a_k} \otimes O_A \otimes O_B) \sigma_{out,j}]
    \end{equation}
    with $O_A$ and $O_B$ generic observables on the two pointers.
    
    
    By defining $\Pi_1=\ket1\bra1=(1-Z)/2$ the projector on the $-1$ eingenstate of $Z$ and $s_{A,B}=\sin(\theta_{A,B})$, $c_{A,B}=\cos(\theta_{A,B})$, $t_{A,B}=\tan
    (\frac{\theta_{A,B}}{2})$, $\Nab=\frac{d}{4\sin\theta_A\sin\theta_B}$, 
    after straightforward calculations we have
    \begin{equation}
    \begin{aligned}
    \langle X_A X_B \rangle_{j,k}&=
    \frac{1}{2\Nab} \left[ (1-\delta_{jk}) \text{Re} (\varrho_{jk}) + \delta_{jk}  \sum_{l\neq j} \text{Re} (\varrho_{jl}) +2c_A \delta_{jk} \text{Re} (\varrho_{jk}) \right] +
    \frac{(c_B-1)}{d\cdot \Nab} \left[ \sum_{l \neq j} \text{Re} (\varrho_{jl}) +c_A \varrho_{jj}  \right]
    \\
    \langle X_A Y_B\rangle_{j,k}&=\frac{1}{2\Nab} \left[\text{Im} (\varrho_{jk})- \delta_{jk}  \sum_{l\neq j} \text{Im} (\varrho_{jl})\right]
    \\
    \langle Y_A X_B \rangle_{j,k}&=\frac{1}{2\Nab} \left[\text{Im}(\varrho_{jk})+\delta_{jk}\sum_{l\neq j}\text{Im}(\varrho_{jl})+\frac{2(c_B-1)}{d}\sum_l\text{Im}(\varrho_{jl})\right]
    \\
    \langle Y_A Y_B\rangle_{j,k}&=\frac{1}{2\Nab} \left[ -\text{Re} (\varrho_{jk}) + \delta_{jk}  \sum_{l} \text{Re} (\varrho_{jl})\right]
    \\
    \langle \Pi_{1A}X_B \rangle_{j,k} &= \delta_{jk} \frac{s_A}{2\Nab} \varrho_{jk} + \frac{s_A(c_B-1)}{2d\cdot \Nab}\varrho_{jj}
    \\
    \langle X_A\Pi_{1B} \rangle_{j,k}&= \frac{s_B}{2d\cdot \Nab} \sum_{l}\text{Re}(\varrho_{jl}) + \frac{s_B(c_A-1)}{2d\cdot \Nab}\varrho_{jj}
    \\
    \langle Y_A\Pi_{1B} \rangle_{j,k}&= \frac{s_B}{2d\cdot\Nab} \sum_{l}\text{Im}(\varrho_{jl}) 
    \\
    \langle \Pi_{1A}\Pi_{1B} \rangle_{j,k}&=\frac{1}{16\Nab^2} \varrho_{jj}
    \end{aligned}
    \end{equation}
    Notice that the last equation is independent of $k$.
    By using the above relations, equations (7) and (8)
    can be easily derived.

\end{document}